\documentclass
[prl,10pt,letterpaper,twocolumn,showpacs,bibnotes,notitlepage,final,superscriptaddress,balancelastpage]{revtex4}%
\usepackage{amssymb}
\usepackage{amsmath}
\usepackage{amsfonts}
\usepackage{graphicx}%
\begin{document}
\title{All-Versus-Nothing Violation of Local Realism by Two-Photon, Four-Dimensional Entanglement}
\author{Tao Yang}
\affiliation{Hefei National Laboratory for Physical Sciences at Microscale and Department
of Modern Physics, University of Science and Technology of China, Hefei, Anhui
230026, China}
\author{Qiang Zhang}
\affiliation{Hefei National Laboratory for Physical Sciences at Microscale and Department
of Modern Physics, University of Science and Technology of China, Hefei, Anhui
230026, China}
\author{Jun Zhang}
\affiliation{Hefei National Laboratory for Physical Sciences at Microscale and Department
of Modern Physics, University of Science and Technology of China, Hefei, Anhui
230026, China}
\author{Juan Yin}
\affiliation{Hefei National Laboratory for Physical Sciences at Microscale and Department
of Modern Physics, University of Science and Technology of China, Hefei, Anhui
230026, China}
\author{Zhi Zhao}
\affiliation{Hefei National Laboratory for Physical Sciences at Microscale and Department
of Modern Physics, University of Science and Technology of China, Hefei, Anhui
230026, China}
\affiliation{Physikalisches Institut, Universit\"{a}t Heidelberg, Philosophenweg 12,
D-69120 Heidelberg, Germany}
\author{Marek \.{Z}ukowski}
\affiliation{Instytut Fizyki Teoretycznej i Astrofizyki Uniwersytet Gda\'{n}ski, PL-80-952
Gda\'{n}sk, Poland}
\author{Zeng-Bing Chen}
\email{zbchen@ustc.edu.cn}
\affiliation{Hefei National Laboratory for Physical Sciences at Microscale and Department
of Modern Physics, University of Science and Technology of China, Hefei, Anhui
230026, China}
\affiliation{Physikalisches Institut, Universit\"{a}t Heidelberg, Philosophenweg 12,
D-69120 Heidelberg, Germany}
\author{Jian-Wei Pan}
\email{pan@ustc.edu.cn}
\affiliation{Hefei National Laboratory for Physical Sciences at Microscale and Department
of Modern Physics, University of Science and Technology of China, Hefei, Anhui
230026, China}
\affiliation{Physikalisches Institut, Universit\"{a}t Heidelberg, Philosophenweg 12,
D-69120 Heidelberg, Germany}
\date{\today }

\begin{abstract}
We develop and exploit a source of two-photon four-dimensional entanglement to
report the first two-particle all-versus-nothing test of local realism with a
linear optics setup, but without resorting to a non-contextuality assumption.
Our experimental results are in well agreement with quantum mechanics while in
extreme contradiction with local realism. Potential applications of our
experiment are briefly discussed.

\end{abstract}

\maketitle

Bell's theorem \cite{Bell} resolves the Einstein-Podolsky-Rosen (EPR) paradox
\cite{EPR}. Arguably, it shows the most radical departure of quantum mechanics
(QM) from classical intuition. It states that certain statistical correlations
predicted by QM for measurements on (originally) two-qubit ensembles cannot be
understood within a realistic picture, based on local properties of each
individual particle. However, Bell's inequalities are not violated by perfect
two-qubit correlations. Strikingly, one also has Bell's theorem without
inequalities for multi-qubit Greenberger-Horne-Zeilinger (GHZ) states
\cite{GHZ,Mermin}. The contradiction between QM and local realism (LR) arises
for definite predictions. LR can thus, in theory, be falsified in a single run
of a certain measurement. This is often called as the \textquotedblleft
all-versus-nothing\textquotedblright\ (AVN)\ proof \cite{Mermin} of Bell's
theorem. Since the GHZ contradiction pertains to definite predictions, and for
all systems, the GHZ theorem represents the strongest conflict between QM and
LR. Further, since it involves perfect correlations, it \emph{directly} shows
that the (based on such correlations) concept of elements of reality, the
missing factor in QM according to EPR, is self-contradictory.

The original GHZ reasoning is for at least three particles and three separated
observers. One may ask: Can the conflict between QM and LR arise for
two-particle systems, for the definite predictions, and for the whole
ensemble? Namely, can the GHZ reasoning be reduced to a two-party (thus two
space-like separated regions) case while its AVN feature is still retained? If
so, one can then refute LR in the simplest and the most essential (i.e.,
unreducible) way. Further, since the EPR reasoning involved only two
particles, such a refutation would be even more direct counterargument against
the EPR ideas than the three-particle one. In a recent exciting debate
\cite{Cabello-Hardy,Cabello-GHZ,Lvovsky,comment,Chen-GHZ} it has been shown
that an AVN violation of LR does exist for two-particle four-dimensional
entangled systems \cite{Chen-GHZ}. In this new refutation of LR, one recovers
EPR's original situation of two-party perfect correlation, but now with much
less complexity. This refutation of LR becomes possible only after introducing
a completely new concept \cite{Chen-GHZ} to define local elements of reality
(LERs). The work in Refs.
\cite{Cabello-Hardy,Cabello-GHZ,Lvovsky,comment,Chen-GHZ} thus demolishes the
original EPR reasoning at the very outset. Here we report the first two-party
AVN test of LR by developing and exploiting a source which produces a
two-photon state entangled both in polarization and in spatial degrees of
freedom.%
\begin{figure}
[ptb]
\begin{center}
\includegraphics[
height=2.2675in,
width=2.6998in
]%
{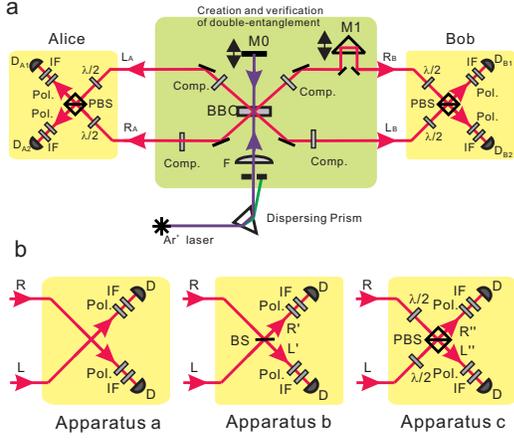}%
\caption{Experimental Setups. (a) An ultraviolet beam from Argon ion laser
(351.1 nm, 120 mW) is directed into the BBO crystal from opposite directions,
thus can create photon pairs (with wave length 702.2 nm) in $\left\vert
\Psi\right\rangle $. Four compensators (Comp.) are used to offset the
birefringent effect caused by the BBO crystal during parametric
down-conversion. The reflection mirrors M0 and M1 are mounted on translation
stages, to balance each arm of the interferometer and to optimize the
entanglement in path. (b) Apparatuses to measure all necessary observables of
doubly entangled states. D is single-photon count module, with collection and
detection efficiency 26\%; IF is interference filter with a bandwidth of
2.88nm and a center wavelength of 702.2nm; Pol. is polarizer. Apparatus c has
been included in (a) at the locations of Alice and Bob.}%
\end{center}
\end{figure}

The experimental setups to generate (Fig. 1a) and to measure (Fig. 1b) pairs
of polarization and path entangled photons are shown in Fig. 1. A pump pulse
passing through a BBO ($\beta$-barium borate) crystal can spontaneously
create, with a small probability, via the parametric down-conversion
\cite{Kwiat}, polarization-entangled photon pairs in the spatial (path) modes
$L_{A}$ and $R_{B}$. For definiteness, we prepare the entangled photon pairs
to be in the maximally entangled state of polarizations $\left\vert \Psi
^{-}\right\rangle _{pol}=\frac{1}{\sqrt{2}}(\left\vert H\right\rangle
_{A}\left\vert V\right\rangle _{B}-\left\vert V\right\rangle _{A}\left\vert
H\right\rangle _{B})$, where $\left\vert H\right\rangle $ ($\left\vert
V\right\rangle $) stands for photons with horizontal (vertical) polarization.
Now if the pump is reflected through the crystal a second time, then there is
another possibility for producing entangled pairs of photons again in
$\left\vert \Psi^{-}\right\rangle _{pol}$, but now into the other two path
modes $R_{A}$ and $L_{B}$. Both pair-creation probabilities can made be equal
by adjusting the foci and location of the focusing lens F. The two possible
ways of producing the (polarization) entangled photon pairs may interfere: If
there is perfect temporal overlap of modes $R_{A}$ and $L_{A}$ and of modes
$R_{B}$ and $L_{B}$, the path state of the pairs is $\left\vert \Psi^{-}%
(\phi)\right\rangle _{path}=\frac{1}{\sqrt{2}}(\left\vert R\right\rangle
_{A}\left\vert L\right\rangle _{B}-e^{i\phi}\left\vert L\right\rangle
_{A}\left\vert R\right\rangle _{B})$, which is also maximally entangled. Here
the two orthonormal kets $\left\vert L\right\rangle $ and $\left\vert
R\right\rangle $ denote the two path states of photons. By properly adjusting
the distance between the mirror and the crystal, so that $\phi=0$, the setup
in Fig. 1a generates the state \cite{Chen-GHZ,Simon-Pan} $\left\vert
\Psi\right\rangle =\left\vert \Psi^{-}\right\rangle _{pol}\otimes\left\vert
\Psi^{-}(0)\right\rangle _{path}$, which is exactly the desired maximally
entangled state in both polarization and path. Actually $\left\vert
\Psi\right\rangle $ can also be interpreted as a maximally entangled state of
two four-dimensional subsystems in a $4\otimes4$ dimensional Hilbert space
\cite{Chen-GHZ}. Figure 2 shows how to achieve good temporal overlap of modes
$R_{A}$ and $L_{A}$ and of modes $R_{B}$ and $L_{B}$ and to adjust the phase
$\phi=0$.%

\begin{figure}
[ptb]
\begin{center}
\includegraphics[
height=1.6912in,
width=3.2559in
]%
{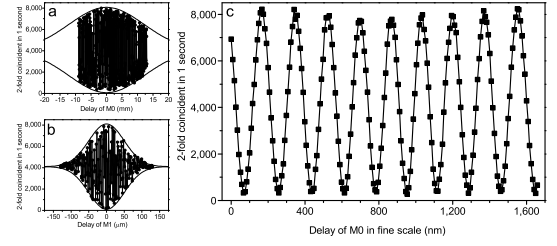}%
\caption{Interference fringe observed when M0 being moved to achieve perfect
temporal overlap and to adjust the phase $\phi=0$. (a,b) Observed interference
fringe for roughly achieving the temporal overlap of modes $R_{A}$ and $L_{A}$
and of modes $R_{B}$ and $L_{B}$. We measure the twofold coincidence between
the output modes toward detectors D$_{A2}$ and D$_{B2}$ behind 45$^{\circ}$
polarizers, by scanning the position of M0 with a step sizes of 1 mm (a) and
of 1 $\mu$m (b). The envelope of the observed twofold coincidence varies
indicating the visibility of the two-photon coherence. Inside the coherent
region, the best visibility is obtained at the position where perfect temporal
overlap is achieved. We perform fine adjustment of the position of M1 and
repeat the scanning of M0 until the best visibility is obtained. (c) We use a
piezo translation stage to move the mirror M0 to perform a fine scan around
the centre of the envelope, with a step size of 1.6 nm. By setting the piezo
system to a position where we observe maximum twofold coincidence of D$_{A2}%
$-D$_{B2}$, we then achieve $\phi=0$.}%
\end{center}
\end{figure}

Then photon-$A$ and photon-$B$ are, respectively, sent to Alice and Bob
(actually the two observation stations are about 1 meter apart in our
experiment). We emphasize that $\left\vert \Psi\right\rangle $ indeed
corresponds to the case where there is one and only one pair production after
the pump pulse passes twice through the BBO crystal. We\textit{ }observed
about $3.2\times10^{4}$ doubly-entangled photon pairs per second.

One can define the following set of local observables to be measured by Alice
and Bob: $\left\vert H\right\rangle \left\langle H\right\vert -\left\vert
V\right\rangle \left\langle V\right\vert \equiv z$ and $\left\vert
+\right\rangle _{pol}\left\langle +\right\vert -\left\vert -\right\rangle
_{pol}\left\langle -\right\vert \equiv x$\ ($\left\vert R\right\rangle
\left\langle R\right\vert -\left\vert L\right\rangle \left\langle L\right\vert
\equiv z^{\prime}$ and $\left\vert +\right\rangle _{path}\left\langle
+\right\vert -\left\vert -\right\rangle _{path}\left\langle -\right\vert
\equiv x^{\prime}$) are two Pauli-type operators for the polarization (path)
degree of freedom of photons. Here $\left\vert \pm\right\rangle _{pol}
=\frac{1}{\sqrt{2}}(\left\vert H\right\rangle \pm\left\vert V\right\rangle )$
and $\left\vert \pm\right\rangle _{path} =\frac{1}{\sqrt{2}}(\left\vert
R\right\rangle \pm\left\vert L\right\rangle )$. Further on, Alice's
observables will be specified by subscript $A$ and Bob's by subscript $B$.

According to Ref. \cite{Chen-GHZ} the six local operators $z_{A}$,
$z_{A}^{\prime}$, $x_{A}$, $x_{A}^{\prime}$, $z_{A}z_{A}^{\prime}$, and
$x_{A}x_{A}^{\prime}$ for Alice ($z_{B}$, $z_{B}^{\prime}$, $x_{B}$,
$x_{B}^{\prime}$ , $z_{B}x_{B}^{\prime}$, and $x_{B}z_{B}^{\prime}$ for Bob)
can be utilized to define the LERs for the two-party system. This is due to
the fact that for the two photons described by $\left\vert \Psi\right\rangle $
QM makes the following predictions:
\begin{gather}
\left.  z_{A}\cdot z_{B}\left\vert \Psi\right\rangle =-\left\vert
\Psi\right\rangle ,\;z_{A}^{\prime}\cdot z_{B}^{\prime}\left\vert
\Psi\right\rangle =-\left\vert \Psi\right\rangle ,\right. \label{re21}\\
\left.  x_{A}\cdot x_{B}\left\vert \Psi\right\rangle =-\left\vert
\Psi\right\rangle ,\;x_{A}^{\prime}\cdot x_{B}^{\prime}\left\vert
\Psi\right\rangle =-\left\vert \Psi\right\rangle ,\right. \label{re22}\\
\left.  z_{A}z_{A}^{\prime}\cdot z_{B}\cdot z_{B}^{\prime}\left\vert
\Psi\right\rangle =\left\vert \Psi\right\rangle ,\ \left.  x_{A}x_{A}^{\prime
}\cdot x_{B}\cdot x_{B}^{\prime}\left\vert \Psi\right\rangle =\left\vert
\Psi\right\rangle ,\right.  \right. \label{re31}\\
\left.  z_{A}\cdot x_{A}^{\prime}\cdot z_{B}x_{B}^{\prime}\left\vert
\Psi\right\rangle =\left\vert \Psi\right\rangle ,\ \left.  x_{A}\cdot
z_{A}^{\prime}\cdot x_{B}z_{B}^{\prime}\left\vert \Psi\right\rangle
=\left\vert \Psi\right\rangle ,\right.  \right. \label{re32}\\
\left.  z_{A}z_{A}^{\prime}\cdot x_{A}x_{A}^{\prime}\cdot z_{B}x_{B}^{\prime
}\cdot x_{B}z_{B}^{\prime}\left\vert \Psi\right\rangle =-\left\vert
\Psi\right\rangle .\right.  \label{re4}%
\end{gather}

Now the nine local variables for Alice will be arranged into three
groups/devices: $a_{A}=(z_{A}^{\prime},x_{A},x_{A}\cdot z_{A}^{\prime})$,
$b_{A}=(z_{A},x_{A}^{\prime},z_{A}\cdot x_{A}^{\prime})$ and $c_{A}%
=(z_{A}z_{A}^{\prime},x_{A}x_{A}^{\prime},z_{A}z_{A}^{\prime}\cdot x_{A}%
x_{A}^{\prime})$, while for Bob $a_{B}=(z_{B},z_{B}^{\prime},z_{B}\cdot
z_{B}^{\prime})$, $b_{B}=(x_{B},x_{B}^{\prime},x_{B}\cdot x_{B}^{\prime})$ and
$c_{B}=(z_{B}x_{B}^{\prime},x_{B}z_{B}^{\prime},z_{B}x_{B}^{\prime}\cdot
x_{B}z_{B}^{\prime})$. For each operational situation (e.g., $a_{A}$) the
observer receives two-bit readouts (results). The bit values, because of the
use of product variables, are here denoted as $\pm1$, instead of $0$ and $1$.
In the case of the device $a_{A}$ Alice can read out $x_{A}$ and
$z_{A}^{\prime}$, and by multiplication get $x_{A}\cdot z_{A}^{\prime}$. With
$b_{A}$ she can measure the values of $z_{A}$ and $x_{A}^{\prime}$ and
therefore fix the derivative value of their product $z_{A}\cdot x_{A}^{\prime
}$. Finally if her choice is $c_{A}$ she gets $z_{A}z_{A}^{\prime}$ and
$x_{A}x_{A}^{\prime}$ and their algebraic product $z_{A}z_{A}^{\prime}\cdot
x_{A}x_{A}^{\prime}$. It is important to note that the last value is not
operationally equivalent to $z_{A}\cdot z_{A}^{\prime}\cdot x_{A}\cdot
x_{A}^{\prime}$, and that it is impossible to measure all these values in the
product for a single system. Similarly, Bob can choose between three
operational situations, namely $a_{B}$ via which he gets the access to $z_{B}$
and $z_{B}^{\prime}$ and their product $z_{B}\cdot z_{B}^{\prime}$, $b_{B}$
which gives $x_{B}$, $x_{B}^{\prime}$ and $x_{B}\cdot x_{B}^{\prime}$, and
finally $c_{B}$ producing $z_{B}x_{B}^{\prime}$, $x_{B}z_{B}^{\prime}$ and
$z_{B}x_{B}^{\prime}\cdot x_{B}z_{B}^{\prime}$. If the above measurements are
performed in spacelike separated regions, then by Einstein's locality, any
measurement performed on one photon would not in any way disturb actions on,
and results for, the other photon. Following EPR, the perfect correlations in
Eqs. (\ref{re21})-(\ref{re4}) allow a local realistic interpretation by
assigning pre-existing measurement values to operators or operator products
that are separated by ($\cdot$). These values \cite{Cabello-GHZ,Chen-GHZ}
would be EPR's elements of reality. Each operational situation for Bob can be
used to establish the EPR elements of reality for three of Alice's variables.
And since we have listed nine perfect correlations, three for each operational
situation at Alice's side, all the above listed variables of Alice seemingly,
according to EPR, can be associated with elements of reality. The same holds
for Bob's variables.

However the above system of LERs turns out to be inconsistent. Let
$m(\Lambda)$ stand for the LER associated with the variable $\Lambda$. If the
quantum perfect correlations are to be reproduced, the following relations
between the LERs must hold:
\begin{gather}
m(z_{A})m(z_{B})=-1,\ m(z_{A}^{\prime})m(z_{B}^{\prime})=-1,\\
m(x_{A})m(x_{B})=-1,\ m(x_{A}^{\prime})m(x_{B}^{\prime})=-1,\\
m(z_{A}z_{A}^{\prime})m(z_{B})m(z_{B}^{\prime})=1,\\
m(x_{A}x_{A}^{\prime})m(x_{B})m(x_{B}^{\prime})=1,\\
m(z_{A})m(x_{A}^{\prime})m(z_{B}x_{B}^{\prime})=1,\\
m(x_{A})m(z_{A}^{\prime})m(x_{B}z_{B}^{\prime})=1,\\
m(z_{A}z_{A}^{\prime})m(x_{A}x_{A}^{\prime})m(z_{B}x_{B}^{\prime})m(x_{B}%
z_{B}^{\prime})=-1.
\end{gather}

Note that the values of, say $z_{A}$ and $z_{A}^{\prime}$, are defined in
different operational situations, namely $b_{A}$ and $a_{A}$, while the value
of the variable $z_{A}z_{A}^{\prime}$ is obtainable operationally in situation
$c_{A}$. This dispels the possible fear of the reader that a local
non-contextuality assumption is tacitly used here - \emph{nowhere do we assume
that $m(z_{A}z_{A}^{\prime})=m(z_{A})m(z_{A}^{\prime})$}, etc. Note that for
all our variables $m(\Lambda)=\pm1$. If one multiplies any subset of the eight
equalities from the above set, side by side, then as a result one gets the
ninth one, but with a wrong sign. That is, if LR holds, and the LERs satisfy
eight of the above relations, which they must, if they are to reproduce the
eight quantum predictions, then on the level of gedanken experiment, LR
predicts that every measurement of the LERs related with the ninth equation
must give a perfect correlation of local results, which, however is perfectly
opposite to the quantum prediction. Outcomes predicted to definitely occur by
LR are never allowed to occur by QM and vice versa. Thus, one indeed has an
AVN conflict between LR and QM.

Importantly, Ref. \cite{Chen-GHZ} also provided a linear optics implementation
of the above experiments, where both Alice and Bob need to measure nine local
variables arranged in three different operational situations: $a_{A}$, $b_{A}$
and $c_{A}$ for Alice and $a_{B}$, $b_{B}$ and $c_{B}$ for Bob. Figure 1b
shows the devices for measuring all the above local observables: Apparatus a
(b) measures the variables in $a_{A}$\ and $a_{B}$ ($b_{A}$\ and $b_{B}$). By
adjusting the polarizers along the two paths, one can measure the polarization
in either $\left\vert H/V\right\rangle $ or $\left\vert \pm\right\rangle
_{pol}$ basis. The measurements in the $\left\vert \pm\right\rangle _{path}$
basis can be achieved by interfering the two paths at a beam splitter (BS)
which affects the transformations $\left\vert R\right\rangle \rightarrow
\left\vert +\right\rangle _{path} $\ and $\left\vert L\right\rangle
\rightarrow\left\vert -\right\rangle _{path} $.

Apparatus c in Fig. 1b measures simultaneously the variables in $c_{A}$\ or
$c_{B}$, where the observables contain always the polarization and the path
information simultaneously. Let us first consider measuring the former. Note
that a polarizing BS (PBS) transmits horizontal and reflects vertical
polarization. If the optical axes of the two half-wave plates ($\lambda/2$;
HWP) in apparatus c are horizontal, the polarizations of the photons will not
be affected after passing through the HWP. Then for Alice's apparatus c, the
outgoing port $R^{\prime\prime}$ ($L^{\prime\prime}$) of the PBS corresponds
to the case of $z_{A}z_{A}^{\prime}=+1$\ ($z_{A}z_{A}^{\prime}=-1$). For
example, an $H$-polarization photon from $L$-path will appear at the
$R^{\prime\prime}$-port of the PBS, with the result $z_{A}=+1$, $z_{A}%
^{\prime}=-1$ and $z_{A}z_{A}^{\prime}=-1$. At the same time, a $V$%
-polarization photon from $R$-path will appear at the same $R^{\prime\prime}%
$-port of the PBS, with the result $z_{A}=-1$, $z_{A}^{\prime}=+1$ and
$z_{A}z_{A}^{\prime}=-1$. Due to the fact that the photon in both path modes
leaves the PBS simultaneously into $R^{\prime\prime}$-path, the information on
whether the photon was transmitted or reflected will be erased if one measures
the photon polarization in the $\left\vert \pm\right\rangle _{pol}$ basis
along the $R^{\prime\prime}$-path. After such an information erasure, one can
find that the case of $x_{A}x_{A}^{\prime}=+1$ ($x_{A}x_{A}^{\prime}=-1$)
corresponds to the photon in $\left\vert +\right\rangle _{pol}$ polarization
($\left\vert -\right\rangle _{pol}$ polarization). Thus, by choosing
appropriate polarizers, apparatus c can then measure the variables in $c_{A}$
simultaneously. For Bob's apparatus c, the only difference stems from the two
HWP, which now affect the transformations $\left\vert H\right\rangle
\rightarrow\left\vert +\right\rangle _{pol} $\ and $\left\vert V\right\rangle
\rightarrow\left\vert -\right\rangle _{pol} $. Following an argument similar
to Alice's apparatus c, one sees that Bob's apparatus c measures\ $z_{B}%
x_{B}^{\prime}$ and $x_{B}z_{B}^{\prime}$ simultaneously and thus also gives
the result of $z_{B}x_{B}^{\prime}\cdot x_{B}z_{B}^{\prime}$\ at the same time.

As we argued above, the operators or operator products separated by $(\cdot)$
can be identified as EPR's elements of reality. The non-contextuality
assumption is not used if the three variables of each group are measured by
one and the same linear optical device \cite{Lvovsky,Chen-GHZ}.

The measured results are consistent with the QM predictions for $\left\vert
\Psi\right\rangle {}$ with high visibilities: $E(z_{A}\cdot z_{B}%
)=-0.98526\pm0.00094$,\ $E(z_{A}^{\prime}\cdot z_{B}^{\prime})=-0.99571\pm
0.00032$,\ $E(x_{A}\cdot x_{B})=-0.98572\pm0.00092$,\ $E(x_{A}^{\prime}\cdot
x_{B}^{\prime})=-0.92999\pm0.00200$,\ $E(z_{A}z_{A}^{\prime}\cdot z_{B}\cdot
z_{B}^{\prime})=0.98538\pm0.00094$, $E(x_{A}x_{A}^{\prime}\cdot x_{B}\cdot
x_{B}^{\prime})=0.88037\pm0.00296$,\ $E(z_{A}\cdot x_{A}^{\prime}\cdot
z_{B}x_{B}^{\prime})=0.90254\pm0.00269$,\ and\ $E(x_{A}\cdot z_{A}^{\prime
}\cdot x_{B}z_{B}^{\prime})=0.98560\pm0.00092$.\ Here the correlation
functions $E(p)=[C(p=+1)-C(p=-1)]/[C(p=+1)+C(p=-1)]$, where $C(p=\pm1)$ are
the counting numbers when the measured variable $p=\pm1$. Each of the above
data was collected within one second by using apparatus a, b or c in Fig. 1b.%
\begin{figure}
[ptb]
\begin{center}
\includegraphics[
height=2.1048in,
width=2.4948in
]%
{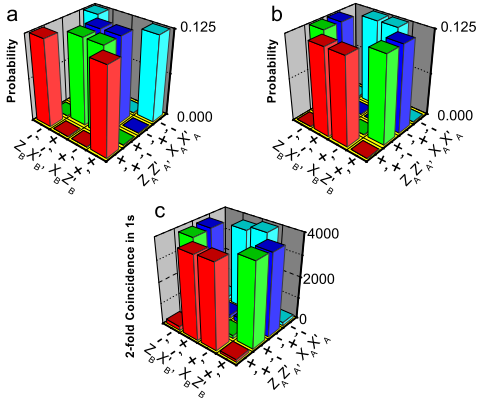}%
\caption{Predictions of LR (a) and of QM (b), and observed results (c) for the
$z_{A}z_{A}^{\prime}\cdot x_{A}x_{A}^{\prime}\cdot z_{B}x_{B}^{\prime}\cdot
x_{B}z_{B}^{\prime}$ experiment.}%
\end{center}
\end{figure}

Once the perfect correlations of $\left\vert \Psi\right\rangle {}$were closely
reproduced in the measurement, we performed the $z_{A}z_{A}^{\prime}\cdot
x_{A}x_{A}^{\prime}\cdot z_{B}x_{B}^{\prime}\cdot x_{B}z_{B}^{\prime}$
($\equiv\mathcal{M}$) experiment, for which QM and LR predict opposite results
(Fig. 3a and 3b). The measured result of $\mathcal{M}$ is shown in Fig. 3c.
With a fidelity of about $96\%$ only those events predicted by QM were
observed in our experiment. This amounts to a very high precision experimental
realization of the first two-photon AVN\ test of LR.

The AVN argument against LR in Ref. \cite{Chen-GHZ} is based on experimentally
unachievable perfect correlations. We also observed spurious events, although
not too often. For instance, although an extremely high fidelity of $96\%$ has
been achieved in the $\mathcal{M}$ experiment, there is still about $4\%$ of
detected events, that is in agreement with LR. Thus, for our argument to hold,
we can assume that these spurious events are only due to experimental
imperfections. Note that, the spurious events are mainly due to the
imperfection of parametric down-conversion source and the limited interference
visibility on the BS and PBS. Alternatively, a Bell-type inequality in Refs.
\cite{Cabello-GHZ,Chen-GHZ} can be used. For any LR model one has
$\left\langle \mathcal{O}\right\rangle _{LRT}\leq7$, where $\mathcal{O}%
=-z_{A}\cdot z_{B}-z_{A}^{\prime}\cdot z_{B}^{\prime}-x_{A}\cdot x_{B}%
-x_{A}^{\prime}\cdot x_{B}^{\prime}+z_{A}z_{A}^{\prime}\cdot z_{B}\cdot
z_{B}^{\prime}+x_{A}x_{A}^{\prime}\cdot x_{B}\cdot x_{B}^{\prime}+z_{A}\cdot
x_{A}^{\prime}\cdot z_{B}x_{B}^{\prime}+x_{A}\cdot z_{A}^{\prime}\cdot
x_{B}z_{B}^{\prime}-\mathcal{M}$. The observed value for $\mathcal{O}$ is
$8.56904\pm0.00533$, which is a violation by about $294$\ standard deviations.

To summarize, with an unprecedented visibility of $95\%$ (i.e. the average of
the absolute value of nine correlation functions observed) we have reported
the first experimental AVN falsification of LR using the two-photon
four-dimensional entanglement. In contrast to previous GHZ experiments
\cite{Pan-GHZ,Zhao-GHZ}, our experiment does not require any post-selection.
This allows an immediate experimental verification of a quantum
pseudo-telepathy game \cite{telepathy}. The high-quality double entanglement
also enables to implement deterministic and highly-efficient quantum
cryptography \cite{Chen-qc} based on the tested AVN falsification of LR. Of
course, as in almost all of the existing experiments testing LR, our
experiment also has certain well known loopholes, such as the locality and
efficiency loopholes. Finally, the full usage of the interference in paths of
photons enables one to entangle two photons in Hilbert space of arbitrarily
high dimensions in a way that is easier than entangling two photons in their
orbital angular momentum states \cite{angular}. Such hyperentanglement and its
manipulation \cite{kwiat98} may be useful in some quantum cryptography
protocols \cite{QC-3level} and in test of Bell's inequalities for
high-dimensional systems \cite{Zeilinger-n}.

The work is supported by the NNSFC and the CAS. This work is also supported by
the Alexander von Humboldt Foundation and the Marie Curie Excellent Grant of
the EU. MZ is supported by Professorial Subsidy of FNP and the MNiL Grant.

\end{document}